\documentclass[accepted=2024-08-16,a4paper,twocolumn,11pt]{quantumarticle}
\pdfoutput=1
\usepackage[utf8]{inputenc}
\usepackage[T1]{fontenc}
\usepackage{amsfonts,amsthm}
\usepackage{bm,graphicx}
\usepackage{bbm}
\usepackage{multirow}
\usepackage[mathscr]{eucal}
\usepackage{xcolor}
\usepackage{amsmath}
\usepackage{dutchcal}
\DeclareMathAlphabet{\dutchcal}{U}{dutchcal}{m}{n}
\usepackage[]{hyperref}
\usepackage{lipsum}
\usepackage[normalem]{ulem}

\usepackage[export]{adjustbox}
\usepackage{tabularray}
\UseTblrLibrary{diagbox}
\usepackage{relsize}

\usepackage[column=O]{cellspace}
\usepackage{tikz}
\usetikzlibrary{decorations.pathreplacing}

\usepackage{ytableau}

\newcommand{\ket}[1]{\left|{#1}\right\rangle}
\newcommand{\bra}[1]{\left\langle{#1}\right|}
\newcommand{\braket}[2]{\langle{#1}|{#2}\rangle}
\newcommand{\ketbrad}[1]{\left|{#1}\rangle\!\langle{#1}\right|}
\newcommand{\ketbra}[2]{\left|{#1}\rangle\langle{#2}\right|}
\newcommand{\mean}[1]{\langle{#1}\rangle}

\newcommand{\tr}{\mathrm{Tr}}

\newmuskip\pFqmuskip

\newcommand*\pFq[6][8]{%
  \begingroup % only local assignments
  \pFqmuskip=#1mu\relax
  \mathchardef\normalcomma=\mathcode`,
  \mathcode`\,=\string"8000
  \begingroup\lccode`\~=`\,
  \lowercase{\endgroup\let~}\pFqcomma
  {}_{#2}F_{#3}{\left(\genfrac..{0pt}{}{#4}{#5}\,;\,#6\right)}%
  \endgroup
}
\newcommand{\pFqcomma}{{\normalcomma}\mskip\pFqmuskip}

\theoremstyle{definition}
\newtheorem{definition}{Definition}

\begin{document}

\title{Quantum multi-anomaly detection}

\author{Santiago Llorens}
\email{santiago.llorens@uab.cat}
\affiliation{F\'{i}sica Te\`{o}rica: Informaci\'{o} i Fen\`{o}mens Qu\`antics, Universitat Aut\`{o}noma de Barcelona, 08193 Bellaterra (Barcelona), Spain}

\author{Gael Sentís}
\email{gael.sentis@uab.cat}
\affiliation{F\'{i}sica Te\`{o}rica: Informaci\'{o} i Fen\`{o}mens Qu\`antics, Universitat Aut\`{o}noma de Barcelona, 08193 Bellaterra (Barcelona), Spain}
\affiliation{Ideaded, Carrer de la Tecnologia, 35, 08840 Viladecans, Barcelona, Spain}

\author{Ramon Mu\~noz-Tapia}
\email{ramon.munoz@uab.cat}
\affiliation{F\'{i}sica Te\`{o}rica: Informaci\'{o} i Fen\`{o}mens Qu\`antics, Universitat Aut\`{o}noma de Barcelona, 08193 Bellaterra (Barcelona), Spain}

\begin{abstract} 
    A source assumed to prepare a specified reference state sometimes prepares an anomalous one. We address the task of identifying these anomalous states in a series of $n$ preparations with $k$ anomalies. We analyze the minimum-error protocol and the zero-error (unambiguous) protocol and obtain closed expressions for the success probability when both reference and anomalous states are known to the observer and anomalies can appear equally likely in any position of the preparation series. We find the solution using results from association schemes theory, thus establishing a connection between graph theory and quantum hypothesis testing. In particular, we use the Johnson association scheme which arises naturally from the Gram matrix of this problem. We also study the regime of large $n$ and obtain the expression of the success probability that is non-vanishing. Finally, we address the case in which the observer is blind to the reference and the anomalous states. This scenario requires a universal protocol for which we prove that in the asymptotic limit, the success probability corresponds to the average of the known state scenario.
\end{abstract}
\maketitle

\section{Introduction.}
Over the past few decades, quantum information theory and its direct application to quantum technologies have advanced rapidly. Numerous quantum tasks in quantum computing~\cite{arute_quantum_2019}, quantum simulation~\cite{georgescu_quantum_2014}, quantum sensing~\cite{degen_quantum_2017}, quantum communications~\cite{khatri_principles_2024}, or quantum machine learning~\cite{dunjko_machine_2018,peral-garcia_systematic_2024} rely on the ability of sources to prepare series of identical quantum states. However, the practical implementation of such identical state preparation faces significant challenges due to noise and imperfections in quantum state generation. In this work, we consider a source that is intended to produce a series of identical states, but it occasionally produces different states. These anomalies can be seen as outliers in a quantum data series, and our task is to identify them. 
This identification can be thought of as an instance of quality control in production, allowing for the application of corrective measures, thereby optimizing resource usage and improving the scalability of quantum systems for technical applications.
This task is also important for the quantum internet \cite{kimble_quantum_2008}, as it can be regarded as a protocol to detect intruders in quantum communications: detecting anomalies in the message may identify which information-carrying system has been compromised.

We study the basic setting of this problem when only a fixed number of mispreparations, or anomalies,  occur. We assume that the anomalous states are known and arbitrary, and it is also promised that a given number of them have been produced, but their positions in the string of prepared states are not known. We aim to certify which of these positions correspond to anomalous states with the minimum error.
Even though our results are derived under these assumptions, they immediately provide a benchmark for multi-anomaly detection strategies in more realistic scenarios, where one should not expect to know the anomalous states, nor these to be pure. Indeed, such generalizations correspond to scenarios where we have less information about the source, hence the success probability of correctly identifying the position of the anomalies will be upper bounded by that of the case of known and pure states.
Partial results for the simplest case of one anomaly were presented in~\cite{skotiniotis_identification_2018}. A somewhat different problem of detecting anomalies from a machine learning perspective was also discussed in~\cite{liu_quantum_2018}.

The problem we are dealing with turns out to be an instance of multi-hypothesis discrimination, for which generically no analytical solutions can be found. Complete solutions are basically only known for the two-state case~\cite{helstrom_quantum_1969}, a result that marks the beginning of the quantum state discrimination field~\cite{chefles_quantum_2000,barnett_quantum_2009,bae_quantum_2015}, and  
sources of symmetric states~\cite{barnett_minimum-error_2001,eldar_optimal_2004}.  Applications to classification problems have recently been addressed~\cite{sentis_unsupervised_2019}, as well as to the detection of Hamiltonian changes~\cite{nakahira_identification_2023}. Asymptotic expressions for some nearly symmetric sources such as the change point~\cite{sentis_quantum_2016} and a few exact zero-error cases are also known~\cite{bergou_optimal_2012,sentis_exact_2017}. 
The multi-anomaly detection task exhibits symmetries that enable us to completely solve the problem.  

This work first aims to derive the optimal solution for the minimum error approach
and to study its asymptotic behavior for a large number of preparations. Additionally, we address the zero error protocol, or unambiguous identification, in which no error is allowed when identifying the anomalies. We also find the optimal universal protocol \cite{fanizza_universal_2022}
that arises when considering that both reference and anomalous states are unknown to the observer. This protocol seeks to provide a robust solution that is applicable across different quantum systems and experimental conditions.

This paper is structured as follows. We first introduce the problem and its Gram formulation and present some mathematical results from graph theory and association schemes that prove to be very suitable for our problem. This toolbox allows us to demonstrate the optimality of the square root measurement (SRM) in a straightforward way. We then present the optimal solution for the minimum error protocol and show that for a fixed number of anomalies $k$, its success probability is non-vanishing in the asymptotic regime of large number of systems $n$.  We next address the unambiguous protocol and compare the results with the minimum error approach. Finally, we present an optimal protocol that is blind to the information of the reference and anomalous states.
We end the paper with three technical appendices, containing a brief account of Johnson graphs and Bose-Mesner algebra, Schur-Weyl duality, and explicit success probability formulae.

\section{Setting of the problem.}
\begin{figure}[!ht]
\centering
  \includegraphics[width=0.8\columnwidth]{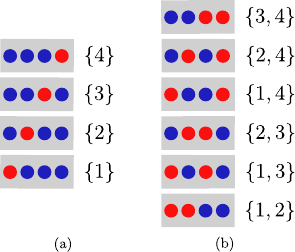}
\caption{Set of hypotheses for the cases of $n=4$.  Subfigure (a)  corresponds to $k=1$, notice that the distance between any two hypotheses is $\delta=1$. Subfigure (b) corresponds to $k=2$ and distances can take values of $\delta=1$, for example hypotheses $\{3,4\}$ and $\{2,4\}$ or $\delta=2$, for instance hypotheses $\{3,4\}$ and $\{1,2\}$.}
\label{fig:hypotheses}
\end{figure}

Let us start by describing the key elements of the multi-anomaly detection problem.
We denote the reference state of the string as $\ket{\phi_0}$ and the anomalous state as $\ket{\phi_1}$. For a given size of the string $n$ and a fixed number $k$ of anomalous states, we have $N= \binom{n}{k}$ possible hypotheses to consider. Without loss of generality, we can consider the overlap between the reference and an anomalous state to be real and denote it as $c=|\braket{\phi_0}{\phi_1}|$. We also assume that each hypothesis has the same prior probability to occur. 
We denote the global state of hypothesis $r$ by  $\ket{\Psi_r}$, with $r$ a subset  of $\{1,\ldots,n\}$ with cardinality $|r|\,=k$ that labels the position of the anomalies (see Fig.~\ref{fig:hypotheses} for simple examples). 

We next construct the  Gram matrix of the set of hypotheses, which contains all the relevant information for a discrimination task. It is defined as the matrix of overlaps of the hypotheses, which in our case reads
\begin{equation}
    G_{rs}=\braket{\Psi_r}{\Psi_s}=(c^2)^{\delta(r,s)}\,,
\end{equation}
where $\delta(r,s):=k-|r\cap s|$ is defined as the distance between two subsets $r$ and $s$, and where $|r\cap s|$ is the number of elements belonging to the intersection of the subsets $r$ and $s$ and corresponds to half the Hamming distance.  In Fig.~(\ref{fig:hypotheses}) we depict the cases of one and two anomalies for $n=4$ systems. In the first case, the distance between different hypotheses is always $\delta=1$, while in the second case, we have $\delta=1$ and $\delta=2$ distances.

Decomposing the Gram matrix $G$ in powers of $c^2$ we obtain a collection of symmetric matrices, that we will denote by $A_i$,  whose entries are 0's and 1's.  Thus, we have
\begin{equation}
\label{eq:gram}
    G=\sum_{i=0}^{k}(c^{2})^i A_i\,.
\end{equation}
In particular, $A_i$ are the adjacency matrices of the generalized Johnson graphs, and form the so-called Bose-Mesner algebra $\dutchcal{A}$ of the Johnson association scheme ~\cite{bannai_algebraic_2021}. The elements of this algebra commute and all have zeros in their diagonal entries, except for the identity matrix $\mathbbm{1}=A_0$ (see Appendix~\ref{app:johnsongraphs} for more details). As we see below, standard results of this association scheme allow us to provide proof of optimality as well as an elegant method to obtain analytical expressions for the success probability.

\section{Minimum error.}
The minimum error protocol consists of a measurement procedure given by a positive operator valued measure (POVM), i.e. a set of positive semidefinite operators $\{\Pi_r\geq 0\}$ that satisfy the completeness relation $\sum_r \Pi_r=\mathbbm{1}$. To each element $\Pi_r$ of the POVM one associates the hypothesis $\ket{\Psi_r}$. The conditional probability of obtaining outcome $\Pi_s$ given a state $\ket{\Psi_r}$ is given by the Born rule $\mathrm{Tr}\left(\ketbra{\Psi_r}{\Psi_r}\Pi_s\right)$ (we have made a slight abuse of notation and labeled the outcome by the same name of the POVM operator). Success occurs when, upon measuring $\ket{\Psi_r}$, one obtains outcome $\Pi_r$ and then the average success probability reads 
\begin{equation}
    P_\mathrm{S}=\sum_{r}\eta_r \text{Tr}\left(\ket{\Psi_r}\bra{\Psi_r}\Pi_r\right),
\end{equation}
where $\eta_r$ is the prior probability of occurrence of hypothesis $\ket{\Psi_r}$. From now onward we refer to the average  $P_\mathrm{S}$ simply as the success probability and assume that all hypotheses have equal priors, $\eta_r=\eta_s=1/N=\binom{n}{k}^{-1}$  $\forall\, r,s$.

The square root $\sqrt{G}=S$ of a Gram matrix defines a measurement, the so-called SRM~\cite{hausladen_pretty_1994}, as the elements $S_{rs}$ are the projections into a measurement basis $\{\ket{m_k}\}$ defined by $S$, i.e., $S_{rs}=\braket{m_r}{\Psi_s}$. Therefore for equal priors, the average success probability of such measurement reads
\begin{equation}
    P_\mathrm{S}=\frac{1}{N}\sum_{r} |S_{rr}|^2\, .
\end{equation}
Now, since the Gram matrix Eq.~\eqref{eq:gram} belongs to a Bose-Mesner algebra $\dutchcal{A}$, any function of $G$ also  
belongs to the same algebra, and in particular its square root $S=\sqrt{G}$.
Therefore, given that the diagonal entries of $G$ are constant, so are the diagonal entries of $S$, that is,  $S_{ll}=S_{hh}=$ $\forall l,h$,  because these only depend on the element of the algebra $A_0=\mathbbm{1}$. Optimality then follows directly, as it is known that for constant diagonal terms of $S$ the SRM is optimal~\cite{sentis_quantum_2016,dalla_pozza_optimality_2015}. 

We next write the success probability in a more convenient way. We note that since the diagonal terms of $S$ are all equal, they can be expressed as $S_{ll}=(\tr S) /N$, with $\tr S =\sum_l \sqrt{\lambda_l}$, where $\lambda_l$  are the eigenvalues of the Gram matrix.
We obtain~\cite{sentis_quantum_2016}
\begin{equation}
    P_\mathrm{S}= \left( \frac{\sum_{l=1}^N \sqrt{\lambda_l}}{N}\right)^2 
    \label{eq:PsSRM}
\end{equation}
Hence, the job reduces to finding the eigenvalues $\lambda_l$.
As every adjacency matrix $A_i$ commutes with each other, they are all simultaneously diagonalizable in some basis, and therefore the eigenvalues of $G$ can be obtained by summing the contributions of a given eigenvalue for each $A_i$ 
\begin{equation}
    \lambda_j(G)=\sum_{i=0}^k (c^2)^i \lambda_j(A_i)\,.
    \label{eq:eigenG_1}
\end{equation}
For the rest of this work, we will omit the dependence of the $G$ eigenvalues and simply write $\lambda_j$ for $\lambda_j(G)$. For the eigenvalues of the adjacency matrices, we will however write the explicit dependence.

From standard results of association schemes the eigenvalues of the first adjacency matrix $A:=A_1$ determine the eigenvalues of any other adjacency matrix of the scheme. The eigenvalue $\lambda_j(A_i)$ is the value of a certain orthogonal polynomial evaluated  at the eigenvalues of $A$ ~\cite{bannai_algebraic_2021}
\begin{equation}
    \lambda_j(A_i)=(-1)^i\binom{k}{i} R_i(\lambda_j(A)+k;0,n-2k,k)\,,
    \label{eq:eigenashahn}
\end{equation}
where $R_i$ is the dual Hahn polynomial of degree $i$~\cite{koekoek_askey-scheme_1996}, which are the orthogonal polynomials arising from the Johnson scheme.
Introducing Eq.~\eqref{eq:eigenashahn} in Eq.~\eqref{eq:eigenG_1} we note that the eigenvalues of $G$ are the generating functions of the dual Hahn polynomials, which read 
\begin{equation}
    \lambda_j=(1-c^2)^{j}\,\pFq{2}{1}{j-k,-n+k+j}{1}{c^2}\,,\label{eq:eigenG_2}
\end{equation}
with multiplicities
\begin{equation}
    m_j=\binom{n}{j}-\binom{n}{j-1}\, ,
    \label{eq:multiplicitieshahn}
\end{equation}
where $j=0,\ldots,k$, and $_{2}F_1$ is the hypergeometric function~\cite{andrews_special_1999} (see Eq.~\eqref{eq:hypergeometric} for further details). Note that the number of distinct eigenvalues is $k+1$, i.e. just the number of anomalies plus one, and are monotonically decreasing with $j$.  For instance, in the simplest case of one anomaly, there are only two different eigenvalues, $\lambda_0=1+(n-1)c^2 >\lambda_1=(1-c^2)$ with multiplicities 1 and $n-1$, respectively.

Taking into account the degeneracy, the success probability of the minimum error protocol finally reads
\begin{equation}
    P_\mathrm{S}=\left(\sum_{j=0}^k \dfrac{m_j}{N}\sqrt{\lambda_j}\right)^2,
     \label{eq:Ps}
\end{equation}
where $\lambda_j$ are given in Eq.~\eqref{eq:eigenG_2}.

\begin{figure}[!ht]
    \centering
    \includegraphics[width=\columnwidth]{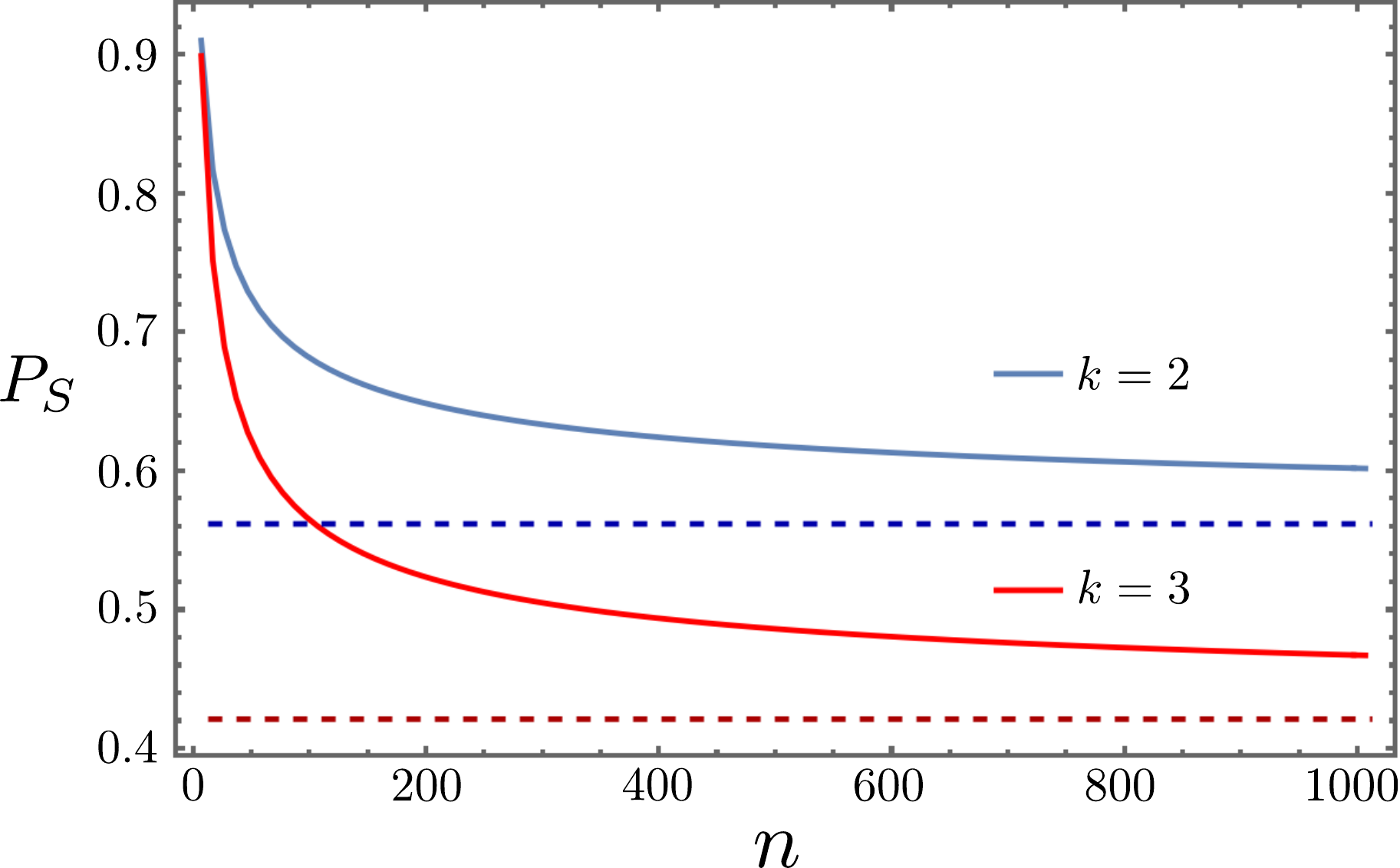}
    \caption{Comparison of the success probability of $2$ anomalies, the upper curve (blue) and the case of $3$ anomalies, the lower curve (red), both for a fixed overlap $c=\frac{1}{2}$. The dashed, upper and lower, lines correspond to the asymptotic limits of the success probabilities of $2$ and $3$ anomalies respectively.}
    \label{fig:multianomalies2and3}
\end{figure}

For a fixed number of anomalies and an increasing number of total systems $n$, it could be expected that the success probability vanishes as the number of possible hypotheses goes to infinity (see Fig.~\ref{fig:multianomalies2and3}). However, the asymptotic limit of $P_\mathrm{S}$ for $n\rightarrow\infty$ is 
finite as we next show.
We note that the eigenvalues of the Gram matrix are polynomials in $n$ of degree $k-j$ ($j=0,1,\ldots,k$), i.e, $\lambda_j\sim O(n^{k-j})$. On the other hand, the ratio between the multiplicities of each eigenvalue $m_j$ and the dimension of the space scales as  $m_j/N\sim O\left(n^{j-k}\right)$ (see Appendix \ref{app:explicit} for the explicit expression of simple cases). Then, it follows that the terms in Eq.~\eqref{eq:Ps} contributing to the first two leading terms are proportional to $\lambda_k$ and  $\sqrt{\lambda_k \lambda_{k-1}}$. 
Substituting the values of the multiplicities, Eq.~\eqref{eq:multiplicitieshahn}, and the explicit expression of the eigenvalues $\lambda_k,\lambda_{k-1}$ given in Eq.~\eqref{eq:eigenashahn} [listed in Table~\eqref{table:eigenG} of Appendix~\ref{app:johnsongraphs}], we finally get
\begin{align}
 P_\mathrm{S}= (1- c^2)^k+ \frac{2kc(1-c^2)^{k-1/2}}{\sqrt{n}}+ O\left(\frac{1}{n}\right)\,.
    \label{eq:psasymptotic}
\end{align}
Note that this expansion is valid in the regime $k/n\ll 1$.

In Fig.~\ref{fig:multianomalies2and3} we depict the success probability for two and three anomalies as a function of the string length.  As can be seen,  the convergence to the asymptotic value is rather slow as the first correction is of order $O(1/\sqrt{n})$. The values of the success probability differ considerably from the asymptotic value, even for a sizeable value of $n$.  This difference is also larger for a larger number of anomalies because of the $k$ factor appearing in the second term of Eq.~\eqref{eq:psasymptotic}.

\section{Zero error identification.}
It is rather straightforward to tackle also the zero error identification scenario, also known as unambiguous discrimination, where one demands that the protocol provides outcomes that identify the position of the anomalies without error. Naturally, this can only happen at the expense of having some inconclusive outcomes. 

The aim is to find a protocol that maximizes the rate of conclusive outcomes, or equivalently minimizes the rate of the inconclusive ones. The POVM has $N+1$ elements: $\{\Pi_r\}$ with $r\subset \{1,\ldots,n\}$ with $|r|=k$, and a $\Pi_0$ element which corresponds to the inconclusive outcome.  Each $\Pi_r$ element identifies with zero error the state $\ket{\Psi_r}$. The structure of the POVM operators is tightly constrained by the zero error conditions $\tr \left(\Pi_r \ketbra{\Psi_s}{\Psi_s}\right)=0$ for $r\neq s$. The optimization can be written as a semidefinite program (SDP)~\cite{boyd_convex_2004,watrous_quantum_2018},
\begin{align}
   \label{eq:primal-ua}
   \begin{split}
   & P_\mathrm{S}  =\frac{1}{N}\max_\Gamma \tr\ \Gamma  \\
   & \mbox{s.t.}\ G-\Gamma_D \geq\ 0 \ \  \mbox{and} \ \ \Gamma \geq 0
    \end{split}   
\end{align}
Here the matrix $\Gamma_D$ is the diagonal part of $\Gamma$ i.e., $[\Gamma_D]_{rs}=\delta_{rs}\Gamma_{rr}$. The entries $\Gamma_{rr}$ correspond to the conditional probabilities that, given hypothesis 
$\ket{\Psi_r}$, it is detected without error.
This form of the SDP is easily derived by noticing that a POVM satisfying the zero error conditions must have elements of the type  $E_k=\gamma_k\ketbrad{\tilde{\varphi}_k}$, where the (unnormalized) states satisfy $\braket{\tilde{\varphi}_k}{\Psi_l}=\delta_{kl}$, i.e., these states are the rows of the (pseudo) inverse of the matrix $ X=\sum_k\ket{\Psi_k}\bra{k}$, where $\ket{k}$ is an orthonormal basis of dimension $N$. The condition $G-\Gamma_D$ follows from multiplying the positivity condition of the inconclusive POVM element, $\mathbbm{1}-\sum_k\gamma_k\ketbrad{\tilde{\varphi}_k}\geq 0$, left and right by the operators $X^\dagger$ and $X$, respectively (see~\cite{sentis_exact_2017} for more details).

To obtain the solution of Eq.~\eqref{eq:primal-ua} we consider the ansatz 
\begin{equation}
\label{eq:mineigenvaule}
    \Gamma=(1-c^2)^k \mathbbm{1}\, ,
\end{equation}
which yields a success probability
\begin{equation}
    \label{eq:ps-ua}
    P_\mathrm{S}=(1-c^2)^k \, .
\end{equation}
We prove that Eq.~\eqref{eq:ps-ua} is the optimal value by checking that (i) $\Gamma$ belongs to the feasibility set (i.e. $G-\Gamma_D \geq\ 0$ and $\Gamma \geq 0$) and that (ii) it induces a feasible instance of the dual SDP in Eq.~\eqref{eq:primal-ua}. Property (i) follows directly from the fact that $(1-c^2)^k=\lambda_{\min}$ (see  Table~\ref{table:dualhahn}).
For the second property let us write the dual program~\cite{sentis_exact_2017}
\begin{align}    
\label{eq:dual-ua}
    \begin{split}     
   & P_\mathrm{S}  =\frac{1}{N} \min_Y\  \tr\ G\, Y\,, \\
     &\mbox{s.t.}\ \ Y_D-\mathbbm{1} \geq\ 0\ \ \mbox{and} \ \ Y \geq 0 \,,
    \end{split}
\end{align}
and let us choose 
\begin{equation}
    Y=\frac{N}{m_k} E_k\,,
\end{equation}
where $E_k$ is the projector onto the minimal eigenvalue of $G$ [see Eq.~\eqref{eq:projectoresE} in Appendix~\ref{app:johnsongraphs}]. From this choice, it follows that $P=\tr\,G\,Y$ coincides with the value of the primal SDP. So, if $Y$  is a feasible solution to Eq.~\eqref{eq:dual-ua}, it is an optimal solution as the values of both primal and dual objective functions coincide. We just have to prove that $Y_D$, the matrix of the diagonal terms of $Y$, satisfies $Y_D\geq \mathbbm{1}$ (note that $Y$ is non-negative by construction as $E_k$ is an orthogonal projection).
Let us decompose $E_k$ as a linear combination of adjacency matrices, $E_k=(1/N) \sum_{j=0}^k q_k(j)A_j$, where $q_k(j)$ are given by the Hahn polynomials [see Eqs~\eqref{eq:E to A} in Appendix~\ref{app:johnsongraphs}] and note again that the only matrix with diagonal elements different from 0 is $A_0=\mathbbm{1}$, i.e.,
\begin{equation}
    Y_D=\frac{q_k(0)}{m_k}\mathbbm{1}\, .
\end{equation}
But $q_k(0)=m_k$ [see Eq.~\eqref{eq:qj} and Table~\ref{table:hahn}] and hence $Y_D=\mathbbm{1}$, which completes the proof.

This result could have been readily anticipated from (i) as the symmetry of the problem already requires that the optimal $\Gamma_D$ satisfies $\Gamma_D \propto\mathbbm{1}$ and then Eq.~\eqref{eq:primal-ua} is just the program defining the lowest eigenvalue of $G$. 
We also see that Eq.~\eqref{eq:ps-ua} coincides with the leading term of the minimum error success probability, Eq.~\eqref{eq:psasymptotic}, hence we conclude that the unambiguous measurement is also optimal for the minimum error protocol for large $n$. This measurement can be realized by a local protocol that checks if each particle has projection or not in the complementary subspace of the reference state. This result can be extended to the case where the anomalies are mixed states since the optimal protocol is fixed by the reference state.

\section{Universal protocol}. We finally tackle the anomaly identification task when the reference and anomalous states are unknown to the observer. That is, we aim at finding a universal protocol that does not use any information about the states (only that they are different, of course) and hence works for an arbitrary pair of reference and anomalous states. In this section, we closely follow the setup and notation of Ref.~\cite{sentis_unsupervised_2019}.

In this scenario, all hypotheses are generated by applying a permutation to the fiducial state, in which all $k$ anomalies are located at the end of the string
\begin{equation}
    \rho_\sigma=\int U_\sigma[\phi_0]^{\otimes (n-k)}\otimes [\phi_1]^{\otimes k}\,U_\sigma^\dagger\, \mathrm{d}\phi_0\, \mathrm{d}\phi_1\, ,
    \label{eq:avg_directions}
\end{equation}
where $[\cdot]:=\ketbrad{\cdot}$, $\phi_{0}$ $(\phi_{1})$ corresponds to the reference (anomalous) state, $U_\sigma$ stands for the unitary transformation that acts on the tensor-product space as a permutation $\sigma\in\mathcal{S}_n\subset S_n$, where $\mathrm{d}\phi_i$ is the measure of the uniform distribution of $\ket{\phi_i}$ and $S_n$ is the symmetric group of $n$ elements. Here, $\mathcal{S}_n$ is the subgroup of relevant permutations for $k$ anomalies, that is, permutations yielding the same state $\rho_\sigma$ are accounted only once.
It is convenient to index each hypothesis by the permutation $\sigma$ that takes a fiducial state where all anomalies occur in the last positions of the hypothesis in question.  Using Schur lemma~\cite{sagan_symmetric_2001,sentis_unsupervised_2019} (see Appendix~\ref{app:schur}), the hypotheses read 
\begin{equation}
    \rho_\sigma=c_k\, U_\sigma\, \mathbbm{1}_{n-k}^{\text{sym}} \otimes \mathbbm{1}_{k}^{\text{sym}} \,U_\sigma^\dagger\,,
    \label{eq:statesunknown}
\end{equation}
where $\mathbbm{1}_k^{\text{sym}}$ refers to the projector onto the fully symmetric subspace of $k$ parties. In the Schur basis, it reads
\begin{equation}
    \rho_\sigma=c_k\,\bigoplus_{\lambda}\mathbbm{1}_\lambda\otimes\Omega_\sigma^\lambda\,,
    \label{eq:hypschur}
\end{equation}
where the normalization constant $c_k$ is 
\begin{equation}
    c_k=\frac{1}{d_k^{\text{\,sym}}d_{n-k}^{\text{\,sym}}}\,,
    \label{eq:normalization}
\end{equation}
and
\begin{equation}
    d_k^{\text{\,sym}}=\binom{k+d-1}{d-1}\,
\end{equation}
is the dimension of the symmetric subspace of $k$ parties.
The parameter $\lambda$ labels the irreducible representation (irrep for short) of the joint action of the groups $\mathrm{SU}(d)$ and the symmetric group $S_n$ over the whole state space $(d,\mathbbm{C})^{\otimes n}$,
 and it is usually identified with bipartitions $(n-\lambda,\lambda)$ of $n$ elements. 
The identity $\mathbbm{1}_\lambda$ is defined over the subspace associated with the irrep $\lambda$ of ${\rm SU}(d)$, and arises in  Eq.~\eqref{eq:hypschur} because we are averaging over all possible states.
The second factor $\Omega_\sigma^\lambda$ stands for an operator that acts in the irrep of $S_n$ labeled by $\lambda$ and it is a rank-1 projector~\cite{sentis_unsupervised_2019}.

We can now derive the explicit expression of the success probability for the minimum-error protocol. The optimal measurement, fortunately, turns out to be again the SRM. For mixed states, its elements are $\Pi_\sigma=\rho^{-1/2}\rho_\sigma\rho^{-1/2}$, where
\begin{equation}
    \rho=\sum_{\sigma\in\mathcal{S}_n} \rho_\sigma=\bigoplus_{\lambda} C_\lambda \mathbbm{1}_\lambda \otimes \mathbbm{1}^\lambda\,.
\end{equation}
where we have used Schur lemma, since $\rho$ commutes with every element $\sigma\in S_n$,
and $C_\lambda$ is a proportionality constant for each partition $\lambda$.
As in previous sections, we write the dual  SDP for the success probability: $\min_Y \tr\, Y$, such that $Y-\rho_\sigma \geq 0$ for all $\rho_\sigma$. If we consider $Y=\sum_\sigma \Pi_\sigma\rho_\sigma$, primal and dual programs give the same value. To prove optimality, one only has to check that $Y$ satisfies the feasibility conditions. Naturally,  $Y-\rho_\sigma\geq 0$, are just the well-known Holevo conditions~\cite{holevo_statistical_1973}.
It is easy to compute that
$
Y=c_k \bigoplus_\lambda \mathbbm{1}_\lambda \otimes \mathbbm{1}^\lambda\,,
$
and to check that  
$
Y-\rho_\sigma\geq0\,, \quad \forall\ \sigma\in\mathcal{S}_n\,,
$
holds, since $\mathbbm{1}^\lambda\geq \Omega_\sigma^\lambda$.
Hence, the optimal success probability  of the universal protocol is just $(1/N) \tr\, Y$, which reads
\begin{align}
    \nonumber P_\mathrm{S}=&\sum_{\lambda=0}^k\Bigg(\frac{(n-2\lambda+1)^2}{(n-\lambda+1)^2}\frac{\binom{n-\lambda+d-1}{d-1}}{\binom{n-k+d-1}{d-1}}\\
    &\qquad\quad\times\frac{\binom{n}{\lambda}}{\binom{n}{k}}\frac{\binom{\lambda+d-2}{d-2}}{\binom{k+d-1}{d-1}}\Bigg)\,,
    \label{eq:psunknown}
\end{align}
where we have used the explicit formulae for the dimensions of the irreps of $\mathrm{SU}(d)$ and $S_n$ for a bipartition $(n-\lambda,\lambda)$ [see Eqs. (\ref{eq:dimsud},\ref{eq:dimsn})].
\begin{figure}[!ht]
\centering
  \includegraphics[width=\columnwidth]{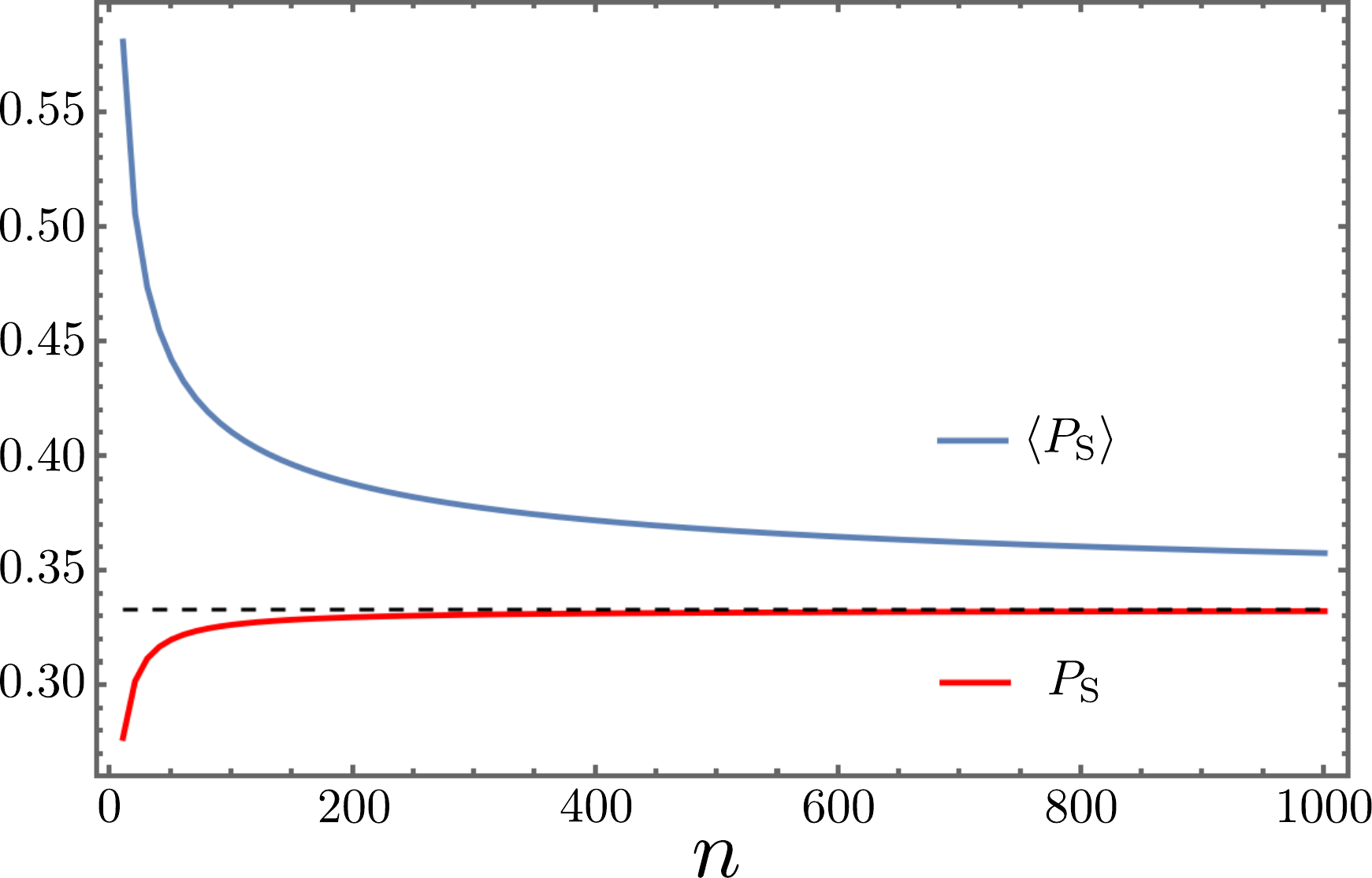}
\caption{Comparison between the average (over $c$) success probability for the case of known reference and anomalous states of dimension $d=2$, the upper curve (blue online), and the success probability for the universal protocol, the lower curve (red online). The dashed black line corresponds to the asymptotic limit of both success probabilities $P_\mathrm{S}=1/(k+1)$} 
\label{fig:avgvsunknown}
\end{figure}

The asymptotic behavior of the success probability in the regime of large $n$ is dominated by the partition $\lambda=k$, as the most important factor in Eq.~\eqref{eq:psunknown} is the ratio $\binom{n}{\lambda}/\binom{n}{k}$  and the contributions of the rest of partitions $\lambda=l<k$ vanish as $O(1/n^{k-l})$. The asymptotic expansion of the success probability then reads
\begin{equation}
    P_\mathrm{S}=\frac{d-1}{d-1+k}+O\left(\frac{1}{n}\right)
\end{equation}
Interestingly, this value coincides with the asymptotic limit of the average success probability for known states and anomalies (see Fig.~\ref{fig:avgvsunknown}),
\begin{align}
    \langle P_\mathrm{S}\rangle\sim & \int_0^1(1-c^2)^k(1-c^2)^{(d-2)}(d-1)\,\mathrm{d}c^2 \nonumber \\
    =& \frac{d-1}{d-1+k}
    \label{eq:mean-ps}
\end{align}
where $(d-1)(1-c^2)^{(d-2)}\mathrm{d}c^2=:\mathrm{d}\mu(c^2)$
is the uniform measure of the square of the overlap~\cite{sentis_unsupervised_2019} and we have
dropped the  vanishing terms in Eq.~\eqref{eq:psasymptotic}.
This agreement has also been observed in the change point problem for unknown states~\cite{llorens_quantum_2024}. The reasons behind this coincidence here stem from the fact that, for known states and large $n$, the optimal measurement approaches an unambiguous protocol that detects anomalies by projecting on the orthogonal space of the reference state. 
Therefore, the knowledge of the anomalous state is not required, for it would be impossible to determine this state with a fixed (and small) number of instances distributed at random places. In this asymptotic limit, only the reference state needs to be known to devise the protocol, and this can be done by using a vanishing number of systems (e.g. $\propto \sqrt{n}$).  The success probability Eq.~\eqref{eq:psunknown} is depicted in Fig.~\ref{fig:avgvsunknown}. We observe that,   as $n$ grows, somehow counterintuitively,  $P_S$ increases.  The logic that explains this behavior is that increasing $n$ allows the protocol to gain more knowledge about the reference state, which gives an advantage that exceeds the detrimental effect of having a larger set of possible hypotheses. It is also evident that, although the universal protocol reaches the value Eq.~\eqref{eq:mean-ps} rather fast,  the average of the success probability for known reference states and anomalies is notably above this value even for significantly large string lengths. Note also that, as expected,  $\mean{P_\mathrm{S}}$ in Eq.~\eqref{eq:mean-ps} tends to one as $d \to \infty$, which reflects the fact that random states tend to be more orthogonal as $d$ grows.

\section{Conclusions.} 
We have done a comprehensive analysis of the problem of identifying anomalous states in a series of allegedly identical quantum state preparations. The accurate identification of faulty preparations will be crucial in the development of real quantum applications. Our results provide a theoretical benchmark of performance under some simplifying assumptions.  A key ingredient to finding closed analytical expressions for the success probability is the theory of association schemes. This mathematical area finds a relevant physical realization in our problem. We have proved that, in the asymptotic limit of large $n$, both optimal minimum error and zero error identification tasks use the same measurement strategy. Furthermore, the success probability is finite and decreases as a power of the number of anomalies. We have also presented a universal protocol that is independent of the states appearing in the problem. Results along these lines have been obtained in the context of quantum learning~\cite{sentis_unsupervised_2019,fanizza_universal_2022} using representation theory and the Schur-Weyl decomposition. 
We have also analyzed the performance of this protocol in the asymptotic limit. We obtain that the success probability coincides with the average of the success probability for known states. This result is quite unexpected in this context since the number of anomalous states is small and these states are located in random places, but it can be explained from the structure of the optimal measurement in this limit.

The tools presented here can be applied to other problems from a novel perspective, e.g., unsupervised classification of states using other figures of merit beyond the error probability.  In general, we expect association schemes such as Hamming and Johnson schemes along with the ones arising from cyclic graphs to be useful for problems in quantum information with appropriate underlying symmetries.

\section{Acknowledgements.}
This work has been financially supported by Ministerio de Ciencia e Innovación of the Spanish Government with funding from European Union Next GenerationEU (PRTR-C17.l1) and by the Generalitat of Catalunya, by the Ministry of Economic Affairs and Digital Transformation of the Spanish Government through the QUANTUM ENIA project: Quantum Spain, by the European Union through the Recovery, Transformation and Resilience Plan - NextGenerationEU within the framework of the “Digital Spain 2026 Agenda”, and from the Spanish Agencia Estatal de Investigación, Grants Nos. PID2019-107609GB-I00 and PID2022-141283NB-I00.
%This work was supported by the Spanish MICIIN with funding from European Union, NextGenerationEU (PRTR-C17.I1), and from the Spanish Agencia Estatal de Investigación, Grants Nos. PID2019-107609GB-I00 and PID2022-141283NB-I00.

\onecolumn
\appendix
\setcounter{secnumdepth}{2}
\newpage
\section{\label{app:johnsongraphs}Johnson scheme}

\begin{figure}[!ht]
    \centering
    \includegraphics[width=0.5\columnwidth]{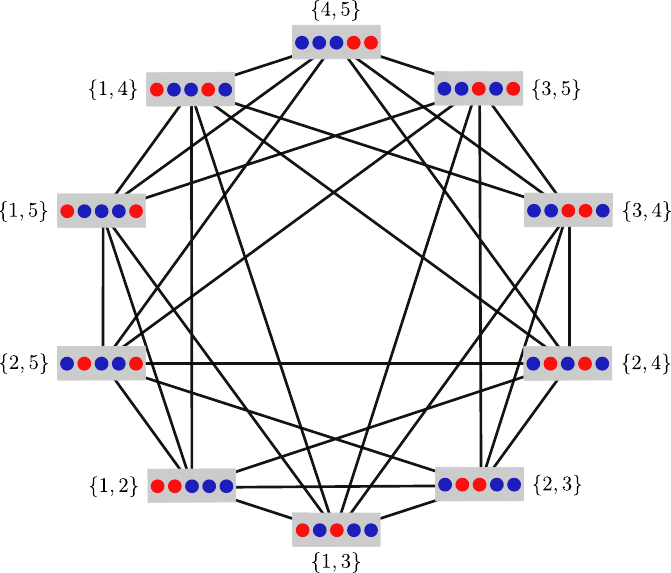}
    \caption{Jonhson graph $J(5,2)$ for a set of $n=5$ objects with subsets of cardinality $k=2$.}
    \label{fig:johnsongraph}
\end{figure}

In this appendix, we introduce the necessary tools of association scheme theory~\cite{bannai_algebraic_2021} for the procedure when solving the multi-anomaly detection problem. First, we will define some necessary concepts, and later we will relate them to our problem. This will allow us to prove optimality conditions for unambiguous discrimination or to compute the values of the success probability of minimum error protocol. 
\begin{definition}
    A Johnson graph $J(n,k)$ is a graph whose vertices are the $k$-element subsets of a $n$-element set. Any pair of vertices $r,s$ are connected by an edge if the intersection of the subsets $|r\cap s|=1$ (see Figure~\ref{fig:johnsongraph}). 
\end{definition}

\begin{definition}
    An adjacency matrix of a graph is a (0,1)-matrix whose entries are defined as 
    \begin{equation}
        A_{rs}=\begin{cases} 
      1 & \text{if vertices } r,s \text{ are adjacent}\,, \\
      0 & \text{otherwise}\,.
   \end{cases}
    \end{equation}
\end{definition}

The $k+1$ eigenvalues and multiplicities of the Johnson graph $J(n,k)$ adjacency matrix are given by
\begin{subequations}
\begin{align}
    \lambda_j(A)=&\ j(n-k-j+1)-k\,,\label{eq:eigenvaluesscheme}\\
    m_j(A)=&\binom{n}{k-j}-\binom{n}{k-j-1}\,,\label{eq:multiplicitiesscheme}
\end{align}
for $j=0,\ldots,k\,$.
\end{subequations}
\begin{definition}
    A generalized Johnson graph $J(n,k,i)$ is a graph whose vertices are the $k$-element subsets of a $n$-element set. Any pair of vertices $r,s$ are connected by an edge if the intersection of the subsets $|r\cap s|=i$. 
    \label{def:johnsongraph}
\end{definition}
\begin{definition}
    A symmetric association scheme is a set of Boolean matrices $\{A_i\}_{i=0}^k$ which satisfy 
    \begin{subequations}
    \begin{align}
        \text{i)}\qquad &\displaystyle\sum_{i=0}^k A_i=J\ \text{(the all-ones matrix)}\,,\\
        \text{ii)}\qquad &A_0=\mathbbm{1}\,,\\
        \text{iii)}\qquad &A_iA_j=A_jA_i=\displaystyle\sum_kp_{ij}^kA_k\,,\label{eq:algebra}
    \end{align}
    \end{subequations}
    for some constants $p_{ij}^k$ called the intersection numbers of the scheme.
\end{definition}
The adjacency matrices of generalized Johnson graphs form the so-called Johnson association scheme. Also, these matrices span a space denoted by $\dutchcal{A}$ of dimension $k+1$, which from Eq.~\eqref{eq:algebra} is closed under multiplication (commutative), that is, it is a commutative algebra that we call Bose-Mesner algebra.

Since all matrices belonging to $\dutchcal{A}$ commute, they are simultaneously diagonalizable in a certain basis, and therefore, $\dutchcal{A}$ has a unique basis of projectors $\{E_i\}_{i=0}^k$ satisfying  
\begin{subequations}
\label{eq:projectoresE}
\begin{align}
        \text{i)}\qquad &E_iE_j=\delta_{ij}E_i\,,\\
        \text{ii)}\qquad &E_0=\frac{1}{N}J\,,\\
        \text{iii)}\qquad &\displaystyle\sum_{i=0}^k E_i=\mathbbm{1}\,,
\end{align}
\end{subequations}
where $N=\binom{n}{k}$, in particular, the adjacency matrices of generalized Johnson graphs can be written on this basis
\begin{equation}
    \label{eq:A to E}
    A_j=\sum_i p_j(i)E_i\,, 
\end{equation}
and vice versa
\begin{equation}
     \label{eq:E to A}
    E_j=\frac{1}{N}\sum_i q_j(i)A_i\,, 
\end{equation}
where, $p_j(i)$ ($q_j(i)$) are the entries of the so-called eigenmatrix $P$ ($Q$) of the scheme, which read
\begin{subequations}
\label{eq:pandq}
\begin{align}
    p_i(j)&=k_i Q_j(i;-n+k-1,-k+1,k)\,,\label{eq:pj}\\
    q_j(i)&=m_j Q_j(i;-n+k-1,-k+1,k)\,,
    \label{eq:qj}
\end{align}
\end{subequations}
where $k_i=\binom{k}{i}\binom{n-k}{i}$ is the valency (number of neighbors) in the Johnson graph $J(n,k,i)$, $m_j$ is the multiplicity of the eigenvalue $j$ in Eq.~\eqref{eq:eigenvaluesscheme} and $Q_j$ stands for the Hahn polynomials~\cite{koekoek_askey-scheme_1996} of degree $j$. In Table~\ref{table:hahn}, we show the first three Hahn polynomials. 
\begin{table}[!h]
    \centering
    \begin{tabular}{|Oc|Oc|Oc|Oc|}
    \hline
        $j$ & $Q_j(x;-n+k-1,-k-1,k)$ \\\hline
        $0$ & $1$ \\\hline
        $1$ & $1-\dfrac{nx}{k(n-k)}$\\\hline
        $2$ & $1-\dfrac{2(n-1)x}{k(n-k)}+\dfrac{(n-1)(n-2)x(x-1)}{k(k-1)(n-k)(n-k-1)}$\\\hline
    \end{tabular}
    \caption{First Hahn polynomials of degree $j$ for arbitrary values of $n$ and $k$.}
    \label{table:hahn}
\end{table}

For a symmetric association scheme, we can write any matrix $A_i$ as a polynomial $v_i(A_1)$, with $A_i$ as the adjacency matrix of the Johnson graph $J(n,k,i)$ and $v_i$ some polynomial of degree $i$. For the Johnson association scheme, we have
\begin{equation}
    A_i=(-1)^i \binom{k}{i} R_i(A+k;0,n-2k,k),
    \label{eq:Aashahn}
\end{equation}
with $R_i$ the dual Hahn polynomial of degree $i$. These polynomials read
\begin{align}
    \nonumber R_i&(\lambda(x);\gamma,\delta,N)\\
    &=\pFq{3}{2}{-i,-x,x+\gamma+\delta+1}{\gamma+1,-M}{1}\,,
    \label{eq:dualhahn}
\end{align}
where $\lambda(x)=x(x+\gamma+\delta+1)$ and the hypergeometric function is defined as 
\begin{equation}
    \pFq{p}{q}{a_1,\ldots,a_p}{b_1,\ldots,b_1}{z}=\displaystyle\sum_{m=0}^\infty \frac{(a_1)_m\ldots(a_p)_m}{(b_1)_m\ldots(b_q)_m}\frac{z^m}{m!}\,,\label{eq:hypergeometric}
\end{equation}
where $(a)_m$ is the rising Pochhammer symbol~\cite{andrews_special_1999}, $(a)_m:=a(a+1)\ldots(a+m)$ and $(a)_0:=1$.
In the Johnson scheme, the values of the variables are given by $x=j=0,\ldots,k$, $\gamma=0$, $\delta=n-2k$ and $M=k$ (see Table~\ref{table:dualhahn} for the values of the first polynomials). Notice from Eqs. (\ref{eq:pj},~\ref{eq:Aashahn}) the duality between the Hahn and the dual Hahn polynomials, as we have written the entries of the eigenmatrix $P$ as the Hahn polynomials, but we could have written them as its dual counterpart, $R_i(\lambda(j);\gamma,\delta,M)=Q_j(i;\gamma,\delta,M)$.

\begin{table}[!ht]
    \centering
    \begin{tabular}{|Oc|Oc|Oc|Oc|}
    \hline
        $i$ & $R_i(\lambda(x);0,n-2k,k)$ \\\hline
        $0$ & $1$ \\\hline
        $1$ & $1-\dfrac{x(n-x+1)}{k(n-k)}$\\\hline
        $2$ & $1-\dfrac{2x(n-x+1)}{k(n-k)}+\dfrac{(x-1)x(n-x+1)(n-x)}{(k-1)k(n-k)(n-k-1)}$\\\hline
    \end{tabular}
    \caption{First dual Hahn polynomials of degree $i$ for arbitrary values of $n$ and $k$.}
    \label{table:dualhahn}
\end{table}

One of the generating functions of the dual Hahn polynomials is given by~\cite{koekoek_askey-scheme_1996}
\begin{align}
    \nonumber&\sum_{i=0}^k(-c^2)^i \binom{k}{i} R_i(\lambda_j+k;0,n-2k,k)\\
    &=(1-c^2)^{k-j}\pFq{2}{1}{-j,-n+2k-j}{1}{c^2}\,,
    \label{eq:generating}
\end{align}
and, as pointed out in the main text, this expression~\eqref{eq:generating}, corresponds to the eigenvalues of the Gram matrix~\eqref{eq:eigenG_1}, which, for the sake of clarity, are presented in Table~\ref{table:eigenG}.   

\begin{table}[ht]
    \centering
    \begin{tabular}{|Oc|Oc|}
    \hline
        $\lambda_k$ & $(1-c^2)^k$ \\\hline
        $\lambda_{k-1}$ & $(1-c^2)^{k-1}[1+c^2(n+1-2k)]$ \\\hline
        $\lambda_{k-2}$ & $(1-c^2)^{k-2}\left[1+\dfrac{1}{2}c^2\left(4+c^2(n+1-2k)\right)(n+2-2k)\right]$\\\hline
        $\vdots$ & $\vdots$ \\\hline
        $\lambda_0$ & $\displaystyle \sum_{i=0}^k  c^{2i} \binom{k}{i}\binom{n-k}{i}$ \\\hline
    \end{tabular}
    \caption{First two smallest and maximum eigenvalues of a Gram matrix with arbitrary parameters $n,k,c$.}
    \label{table:eigenG}
\end{table}

\section{\label{app:schur}Schur-Weyl duality}

In this appendix, we recall a brief introduction to representation theory used for the universal protocol of multi-anomaly detection with a connection of representation theory with the Johnson scheme. 

Schur-Weyl duality establishes a connection between the irreducible representations of the group of linear transformations $\mathrm{GL}(d)$, in particular for our problem, $\mathrm{SU}(d)$, and the symmetric group $S_n$. The action of any transformation $R^{\otimes n}$ with $R\in \mathrm{SU}(d)$, commutes with the action of any $U^\sigma$, the permutation of the tensor-product space $(d,\mathbbm{C})^{\otimes n }$, with $\sigma\in S_n$. Both of these transformations induce a reducible representation over the $\mathrm{SU}(d)$ and $S_n$ groups, and it follows that this representation reduces into irreps. as 
\begin{equation}
    U_\sigma R^{\otimes n}=R^{\otimes n} U_\sigma = \bigoplus_\lambda R_\lambda \otimes U_\sigma^\lambda\,,
\end{equation}
where $\lambda$ is a partition of the tensor-product space $\lambda=(\lambda_1,\lambda_2,\ldots,\lambda_s,\ldots)$, with $\sum_r \lambda_r =n$, that labels both, the irreducible representations of $R^{\otimes n}$ and $U_\sigma$. Note that our notation uses a subscript $\lambda$ when referring to the representation of the $\mathrm{SU}(d)$ group and a superscript when referring to $S_n$. 

This block-diagonal structure allows us to decompose the total Hilbert space $\mathcal{H}^{\otimes n}=(d,\mathbbm{C})^{\otimes n}$ as a direct sum of subspaces $\mathcal{H}^{\otimes n}=\bigoplus_\lambda H_\lambda\otimes H^\lambda$ that are invariant under the action of the groups $\mathrm{SU}(d)$ and $S_n$. The basis in which the Hilbert space has this form is called the Schur basis, and it happens to be very suitable for universal protocols in which no information about the representation of the states in $\mathrm{SU}(d)$ is available. 

In our problem, we make use of this duality, since the only available information contained in our hypotheses is given by the representation of the $S_n$ group. We start from Eq.~\eqref{eq:avg_directions}
where we average over all possible uniform distributions of  $\phi_0$ and  $\phi_1$.  
Then, using Schur lemma~\cite{sagan_symmetric_2001,sentis_unsupervised_2019}, the hypotheses read 
\begin{equation}
    \rho_\sigma=c_k\, U_\sigma\, \mathbbm{1}_{n-k}^{\text{sym}} \otimes \mathbbm{1}_{k}^{\text{sym}} \,U_\sigma^\dagger\,,
    \label{eq:statesunknown_app}
\end{equation}
where $\mathbbm{1}_k^{\text{sym}}$ refers to the symmetric projector onto $k$ parties,  
The normalization constant $c_k$ in Eq.~\eqref{eq:normalization} is the inverse of the dimensions of the symmetric subspaces of $k$ and $n-k$ parties 
\begin{equation}
    c_k=\frac{1}{d_k^{\text{\,sym}}d_{n-k}^{\text{\,sym}}}\,.
\end{equation}
The dimension of the symmetric subspace of $n$ parties corresponds to the dimension of the irrep. of $\mathrm{SU}(d)$ of a bipartition $\lambda=(n,0)$
\begin{equation}
    d_k^{\text{\,sym}}=s_{(k,0)}\,.
\end{equation}
the expression for $s_{(\lambda_1,\lambda_2)}$ is given explicitely in Eq.~\eqref{eq:dimsud}.

States in~\eqref{eq:statesunknown} written using Schur basis~\cite{sagan_symmetric_2001} have a final block-diagonal form of Eq.~\eqref{eq:hypschur}. Notice that $\mathbbm{1}_\lambda$ appears as the irrep. of $\mathrm{SU}(d)$ in all bipartitions $\lambda$, since there is no information left of the possible directions that both reference and anomalous states can have.

Note that since, we are considering only two different states, the reference and the anomalous, the partitions of the tensor-product space will be 2-partitions, i.e. $\lambda=(\lambda_1,\lambda_2)$, with $\lambda_1+\lambda_2=n$. With this at hand, the dimensions of the irreps. $\mathrm{SU}(d)$ and $S_n$ used in Eq.~\eqref{eq:psunknown} have a rather simple expression
\begin{subequations}
\label{eq:dimandmult}
\begin{align}
    s_\lambda&=\frac{\lambda_1-\lambda_2+1}{\lambda_1+1}\binom{\lambda_1+d-1}{d-1}\binom{\lambda_2+d-2}{d-2}\,,\label{eq:dimsud}\\
    m_\lambda&=\binom{n}{\lambda_2}-\binom{n}{\lambda_2-1}\,,\label{eq:dimsn}
\end{align}
\end{subequations}
notice that the label $m$ in Eq.~\eqref{eq:dimsn} for the dimension of the $S_n$ irrep. is not coincidental, as it corresponds to the multiplicity of the eigenvalues of the Jonshon graph in Eq.~\eqref{eq:multiplicitiesscheme}.

\section{Success probability for 1, 2, and 3 anomalies}
\label{app:explicit}

We present the explicit expressions of the success probability for minimum error $P_\mathrm{S}$ for the cases of $k=1,2,3$ anomalies.

$$k=1$$

\begin{align}
    &P_\mathrm{S}^{(1)}=\frac{1}{n^2}\left\{\left(n-1\right)\left(1-c^2\right)^{1/2}+\sqrt{1+(n-1)c^2}\right\}^2\,.
\end{align}

$$k=2$$

\begin{align}
    P_\mathrm{S}^{(2)}=\left\{\frac{n-3}{n-1}\left(1-c^2\right)+\frac{2}{n}(1-c^2)^{1/2}\sqrt{1+(n-3)c^2}
    +\frac{2}{n\left(n-1\right)}\sqrt{1+2\binom{n-2}{1}c^2+\binom{n-3}{2}c^4}\right\}^2\,.
\end{align}

$$k=3$$

\begin{align}
     P_\mathrm{S}^{(3)} & =\left\{\frac{n-5}{n-3}\left(1-c^2\right)^{3/2}+\frac{3\left(n-3\right)}{\left(n-1\right)\left(n-2\right)}(1-c^2) \sqrt{1+c^2(n-5)} \right.
    \nonumber \\
    & + \frac{6}{n (n-2)} 
     (1-c^2)^{1/2}\sqrt{1+2c^2\binom{n-4}{1}+c^2\binom{n-5}{2}}  \nonumber \\
    &+ \left. \frac{6}{n\left(n-1\right)\left(n-2\right)}
     \sqrt{1+3c^2\binom{n-3}{1}+3c^4\binom{n-4}{2}+c^6\binom{n-5}{3}}\right\}^2\, .
\end{align}
\bibliography{Multianomaly}

\end{document}